# Upper critical magnetic field and multiband superconductivity in artificial high-T$_c$ superlattices of nano quantum wells

Gaetano Campi[1,2], Andrea Alimenti[3,4], Gennady Logvenov[5], G. Alexander Smith[6], Fedor. F. Balakirev[6], Sang-Eon Lee[7], Luis Balicas[7], Enrico Silva[3,4], Giovanni Alberto Ummarino[8], Giovanni Midei[9,10], Andrea Perali[11,2], Antonio Valletta[12], Antonio Bianconi[1,2]

[1] Institute of Crystallography, National Research Council, CNR, Via Salaria Km 29.3, 00015 Monterotondo Rome, Italy
[2] Rome International Center for Materials Science Superstripes RICMASS, Via dei Sabelli 119A, 00185 Rome, Italy
[3] Department of Industrial, Electronic and Mechanical Engineering, Roma Tre University, Via Vito Volterra 62, 00146 Rome, Italy
[4] Istituto Nazionale di Fisica Nucleare INFN, Sezione Roma Tre, Via della Vasca Navale 84, 00146 Rome, Italy
[5] Max Planck Institute for Solid State Research, Heisenbergstraße 1, 70569 Stuttgart, Germany
[6] National High Magnetic Field Laboratory, (NHMFL) Los Alamos National Laboratory (LANL), Los Alamos, NM 87545, USA
[7] National High Magnetic Field Laboratory (NHMFL), Florida State University (FSU), Tallahassee, FL 32310, USA.
[8] Istituto di Ingegneria e Fisica dei Materiali, Dipartimento di Scienza Applicata e Tecnologia, Politecnico di Torino, Corso Duca degli Abruzzi 24, 10129 Torino, Italy
[9] School of Science and Technology, Physics Division, University of Camerino, 62032 Camerino, Italy
[10] Istituto Nazionale di Fisica Nucleare, Sezione di Perugia, Via A. Pascoli, 23c, 06123 Perugia, Italy
[11] School of Pharmacy, Physics Unit, University of Camerino, 62032 Camerino, Italy
[12] Institute for Microelectronics and Microsystems, National Reesearch Council CNR, Via del Fosso del Cavaliere 100, 00133 Roma, Italy

- "Gaetano Campi" *email:* gaetano.campi@cnr.it to whom correspondence should be addressed
- "Andrea Alimenti" *email:* andrea.alimenti@uniroma3.it to whom correspondence should be addressed
- "Gennady Logvenov" *email:* g.logvenov@fkf.mpg.de
- "Luis Balicas" *email:* balicas@magnet.fsu.edu
- "Sang-Eon Lee" *email:* sangeon.lee@fsu.edu
- "Smith Gregory Alexander Tilman" *email*: gasmith@lanl.gov
- "Balakirev Fedor Fedorovich" *email*: fedor@lanl.gov
- "Enrico Silva" *email*: enrico.silva@uniroma3.it
- "Giovanni A. Ummarino" *email:* giovanni.ummarino@polito.it
- "Giovanni Midei" *email:* giovanni.midei@unicam.it
- "Antonio Valletta" email: antonio.valletta@cnr.it
- "Andrea Perali" *email:* andrea.perali@unicam.it
- "Antonio Bianconi" *email:* antonio.bianconi@ricmass.eu to whom correspondence should be addressed

Gaetano Campi           https://orcid.org/0000-0001-9845-9394
Andrea Alimenti         https://orcid.org/0000-0002-4459-6147
Gennady Logvenov        https://orcid.org/0000-0003-1986-0249
G. Alexander Smith      https://orcid.org/0000-0001-9701-1328
Fedor. F. Balakirev     https://orcid.org/0000-0003-4887-5140
Sang-Eon Lee            https://orcid.org/0000-0002-7827-6117
Luis Balicas            https://orcid.org/0000-0002-5209-0293
Enrico Silva            https://orcid.org/0000-0001-8633-4295
Giovanni Ummarino       https://orcid.org/0000-0002-6226-8518
Giovanni Midei          https://orcid.org/0009-0008-8970-291X
Andrea Perali           https://orcid.org/0000-0002-4914-4975
Antonio Valletta        https://orcid.org/0000-0002-3901-9230
Antonio Bianconi        https://orcid.org/0000-0001-9795-3913

* to whom correspondence should be addressed:
*email:* gaetano.campi@cnr.it
*email:* andrea.alimenti@uniroma3.it
*email*: antonio.bianconi@ricmass.eu

**Abstract**

Artificial high-$T_c$ superlattices (AHTS) composed of quantum building blocks with tunable superconducting critical temperature have been synthesized by engineering their nanoscale geometry using the Bianconi-Perali-Valletta (BPV) two gaps superconductivity theory. These quantum heterostructures consist of quantum wells made of superconducting, modulation-doped Mott insulators (S), confined by a metallic (N) potential barrier. The lattice geometry has been carefully engineered to induce the predicted Fano-Feshbach shape resonance between the gaps, near a topological Lifshitz transition. Here, we validate the BPV theory by providing compelling experimental evidence that AHTS samples, at the peak of the superconducting dome, exhibit resonant two-band, two-gap superconductivity. This is demonstrated by measuring the temperature dependence of the upper critical magnetic field, $\mu_0 H_{c2}$, in samples with superlattice periods $3.3 < d < 5.28$ nm and L/d ratios close to the magic value 2/3 (where L is the thickness of the superconducting $La_2CuO_4$ layer and d is the superlattice period). The data reveal the predicted upward concavity in $H_{c2}(T)$ and a characteristic kink in the coherence length as a function of temperature, confirming the predicted two-band superconductivity with Fermi velocity ratio $\approx 0.25$ and significant pair exchange term among the two condensates.

*Introduction* - For 29 years, the mechanism of high-$T_c$ superconductivity in cuprate perovskites has remained elusive. Here, we validate a proposed paradigm in which $T_c$ amplification is driven by a Fano-Feshbach resonance controlled by the quantum geometry of the nanoscale building blocks. The two-dimensional electron gas (2DEG) formed at the interface between the insulating dielectric perovskite oxides $LaAlO_3$ and $SrTiO_3$ (LAO/STO) [1,2] demonstrates the realization of a confined 2DEG exhibiting two-band superconductivity [3,4,5] with a low critical temperature. The Bianconi-Perali-Valletta (BPV) theory [6-10] which incorporates quantum exchange interaction for pair transfer between two bands with different gaps has been employed to engineer quantum confined high-temperature superconductivity interfaced between chemically undoped and doped copper oxide [11,12]. The growth of these nanoscale unconventional heterostructures called Artificial High-$T_c$ superlattices (AHTS) was obtained using oxide molecular beam epitaxy [12] guided by the nanomaterials design of variable quantum geometry. The artificially modulation-doped [13] Mott insulator superlattices in the 2-4 nm range [14-17] have confirmed the predictions of the BPV theory including spin orbit coupling which describes the amplification of the critical temperature driven by Fano-Feshbach shape resonances in two-gap interface superconductivity [18-19]. The spin-orbit coupling (SOC) driven by internal contact electric at the interfaces in nanostructured materials exhibiting quantum size effects plays a key role in nanomaterial quantum design of the three-dimensional AHTS [14-17].

Here, we focus on AHTS heterostructures composed of superconducting layers of stoichiometric modulation-doped Mott insulator $La_2CuO_4$ (LCO) with thicknesses $L$, intercalated with potential barriers of normal metal made of chemically overdoped non-superconducting $La_{1.55}Sr_{0.45}CuO_4$ (LSCO), with ten repetitions of the period $d$ whose value remains in the nanoscale range $2.97 < d < 5.28$ nm. The normal metal units act as charge reservoirs, transferring interface space charge into the superconducting Mott insulator units of thickness $L$, thereby forming a 3D superlattice of modulation doped 2D Mott Insulator-Metal Interface (MIMI) with period $d$. The internal interface electric field at the SNS junctions induces Rashba spin-orbit coupling (SOC) in the superconducting interface space charge within the Mott insulator layers, which is split into two electronic subbands by quantum size effects. The lowest subband shows a large cylindrical Fermi surface with high Fermi velocity, while the upper subband exhibits a low Fermi velocity and an unconventional extended van Hove singularity with low group velocity at the Fermi level generated by the SOC at the interface. According to the BPV theory, the critical temperature, $T_c$, is a function of the geometrical parameter $L/d$, reaches a maximum at the magic ratio $L/d = 2/3$ [14-17]. By tuning $L/d$ it is possible to change the energy splitting between the bottom of the two subbands bringing the system at the optimal distance from the electronic topological Lifshitz transition, where the second subband disappears, to reach the maximum of the quantum resonance.

In this study, we investigated the resistivity behavior of AHTS under externally applied magnetic fields ranging from 1 to 58 Tesla. Our experimental data was meticulously gathered at three specialized facilities, ensuring the robustness and reliability of our findings. The primary focus of our study was on superlattices with $L/d$ ratios close to the critical value 2/3. At this critical ratio ($L/d = 2/3$), we observed a significant enhancement in the superconducting properties, due to the $T_c$ amplification at the Fano-Feshbach shape resonance, [17-19] demonstrating that the interplay between the quantum properties of the superlattices and nanoscale structural parameters is well-controlled.

Here, our primary goal is the experimental validation of the theoretical prediction of two-band superconductivity with two distinct Fermi velocities at the peak of the superconducting dome in the AHTS. This is achieved by measuring the temperature dependent upper critical magnetic field $\mu_0 H_{c2}(T)$. It is well known that in type-ll superconductors the $\mu_0 H_{c2}(T)$ curve predicted by the BCS single-band mean field theory, follows the Werthamer-Helfand-Hohenberg (WHH) model [20] given by a simple downward convex function of the reduced $h_{c2}(t)=1-t^2$, where $t=T/T_c$ and $h_{c2}(t)=H_{c2}(T)/H_{c2}(T=0)$. Deviations of the experimental upper critical field $h_{c2}(t)$ from the WHH curve was used as a signature for multi-band and multi-gap superconductivity in bulk $MgB_2$ [21,22] and interfacial STO, [23-25]. Edge and Balatsky [23] have shown theoretically that the upward concavity as a function of the temperature of the upper critical magnetic field $h_{c2}(t)$ is a quantitative probe for multiband superconductivity. This finding has been confirmed by recent experimental and theoretical studies of a large variety of two-band and two-gap superconductors [26-46] and it is further corroborated by our theoretical modeling, which aligns well with the experimental observations. Additionally, we found that the $L/d$ ratio influences scattering phenomena in the normal state. The suppression of superconductivity by high magnetic fields enables the study of the normal state in regions where the superconducting condensate typically obscures the detection of particularly low Kondo temperatures [16]. Here, by suppressing superconductivity, we measured a resistance upturn, providing evidence for the Kondo proximity effect in the normal phase of Mott Insulator Metal Interfaces (MIMI) superlattices.

*Results* - We investigated the resistivity as a function of the temperature in nanoscale artificial high $T_c$ superlattices composed of Mott Insulator Metal Interfaces. The superlattices were formed by alternating layers of $La_2CuO_4$ (LCO) and $La_{1.55}Sr_{0.45}CuO_4$ (LSCO), with the thickness ratio $L/d$ near the magic value of 2/3 [16,17]. Fig. 1(a) illustrates the MIMI structure for the three samples studied in this work, highlighting the LCO (blue) and LSCO (orange) layers. Fig. 1(b) shows the superconducting transition temperatures $T_c$ for the samples having different $L/d$ values, compared with theoretical predictions from the BPV theory. The sheet resistance measurements for the three samples in Fig. 1(c) shows that as $L/d$ approaches the critical ratio of 2/3, the resistivity tends toward a linear regime as discussed in [16]. In Fig. 1(d), we demonstrate that at the peak of the superconducting dome, the normal phase exhibits Planckian $T$-linear resistivity which is proposed to compete with the generalized Kondo scattering effect [17]. Since at lower temperatures, the resistance tends to saturate, to ensure that the low-temperature resistance behavior provides evidence for the Kondo effect, it was necessary to suppress superconductivity. We have achieved this by applying a high out-of-plane magnetic field, $\mu_0H_{//c}$, of 35 T, which reduced the critical temperature $T_c$ from 43 K to 5 K. Under these conditions, we observed the Kondo effect (blue dotted line) on the LCO/LSCO Mott Insulator-Metal Interface with $L/d = 2/3$, alongside the low-temperature behavior of the normal phase (red line).

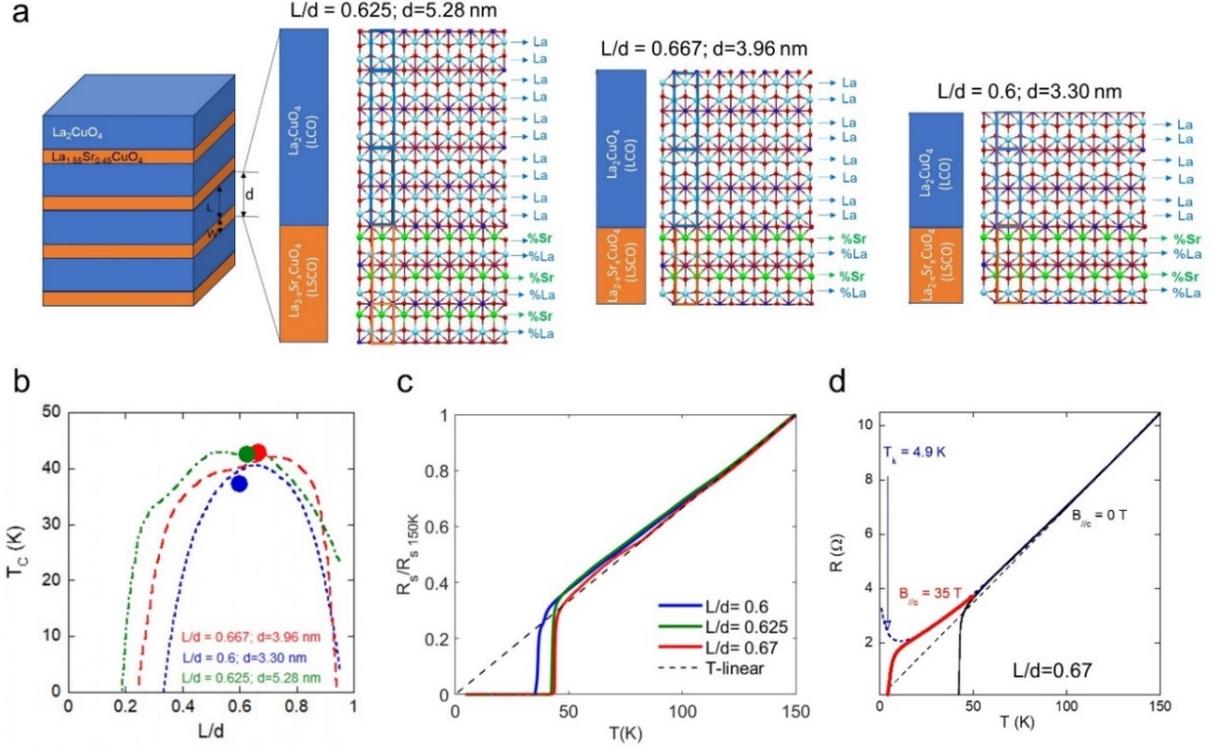

**Figure 1 (a)** Structure of all three samples investigated in this work formed by the Mott insulator $La_2CuO_4$ (blue layers) with thickness $L$, intercalated with potential barriers of normal metal made of chemically overdoped $La_{1.55}Sr_{0.45}CuO_4$ (orange layers). **(b)** Superconducting critical temperature measured in these three samples with different $L/d$ values alongside the theoretical curves calculated by the BPV theory. **(c)** Sheet resistance measured in all three samples. **(d)** Kondo modeling (blue dotted line), while suppressing $T_c$ from 43 K to 5 K, provides the best fitting curve for the resistance as a function of the temperature, measured both with (red line) and without (black line) a magnetic field. Measurements under high magnetic fields were performed at the National High Magnetic Field Laboratory (NHMFL) in Tallahassee. The Kondo temperature, $T_K$, is 5 K.

Modeling of the resistivity in the normal phase between 10 K and 150 K using the universal Kondo equation for MIMI provides the best fit for the measured resistance as a function of the temperature, both with and without a magnetic field, yielding a Kondo temperature $T_K$ of 5 K. In these heterostructures, the Kondo effect is driven by spin-orbit coupling induced by the potential drop at the interface and by the coexistence of itinerant and localized electronic components.

Resistance measurements under high magnetic fields were conducted at three facilities: Roma Tre, MagLab DC Field Facility in Tallahassee, and MagLab Pulsed Field Facility at LANL. Figure 2 displays the resistance data for two samples with specified $L/d$ and $d$ values across different temperature and magnetic field ranges. The measurements from each facility cover distinct $T$ and $\mu_0 H$ ranges, providing a comprehensive view of the resistive behavior under a wide range of externally applied magnetic fields from 1 to 58 tesla.

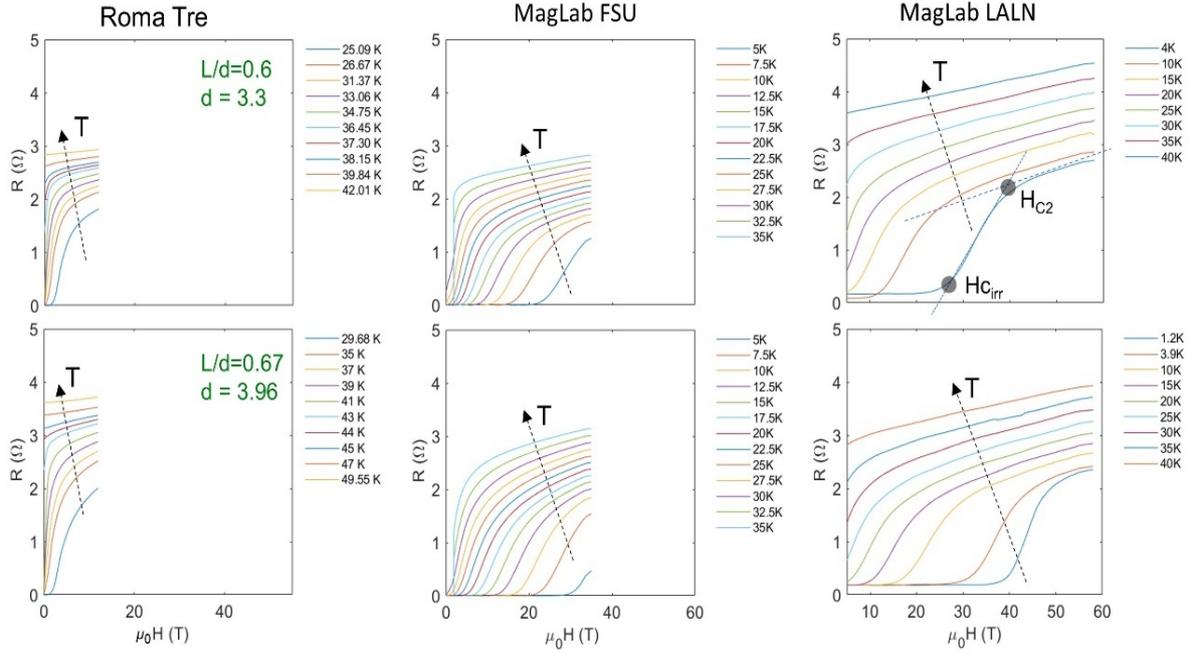

**Figure 2** Resistance $R$ as a function of the magnetic field $\mu_0H$ measured in two samples with different $L/d$ and $d$ values, under high magnetic fields at the three facilities: Roma Tre, MagLab at FSU, and MagLab at LANL. Measurements in each facility cover a specific range of both temperature, $T$, and magnetic field. The $\mu_0H_{c2}$ and $\mu_0H_{Cirr}$ values for the $R(\Omega)$ curve measured at 40 K was extracted as the point intersecting the lines tangent to the curve in the initial and middle part ($\mu_0H_{Cirr}$) and in the middle and final part ($\mu_0H_{c2}$) of the curve. This geometrical procedure was applied to get $\mu_0H_{Cirr}$ and $\mu_0H_{c2}$ values for each $R(\Omega)$ line.

The $\mu_0H_{c2}$ and $\mu_0H_{Cirr}$ values for the resistance $R(\Omega)$ curves measured as a function of $\mu_0H$ at different temperatures were determined first using a geometrical method. This method involves identifying the points where the tangents to the curve intersect. Specifically, an effective experimental $\mu_0H_{Cirr}$ could be extracted at the intersection of the tangents to the initial and middle parts of the curve, while the upper critical field $\mu_0H_{c2}$ is extracted at the intersection of the tangents to the middle and final parts of the curve. This geometrical procedure, illustrated for the line at $T=40K$ in the upper left panel of Fig. 2, was consistently applied to determine the $\mu_0H_{c2}$ and $\mu_0H_{Cirr}$ values for each $R(\Omega)$ curve measured.

The $\mu_0H_{c2}$ and $\mu_0H_{Cirr}$ values were also determined by fitting the derivative of each resistance $R(\Omega)$ curve with respect to the magnetic field ($\mu_0H$) using the asymmetric Fano line-shape:

$$y_{Fano}(H) = b + K \left[\frac{A_s(H-H_{max})}{0.5*\sigma}\right]^2 \Big/ \left[\frac{(H-H_{max})}{0.5*\sigma}\right]^2 \qquad (1)$$

where $\mu_0H_{max}$ is the value of the magnetic field at which the derivative $dR/d(\mu_0H)$ is maximum. $A_s$ and $\sigma$ represent the asymmetry term and the shape width, while $b$ and $K$ are two constants. In this context $\mu_0H_{c2} = \mu_0H_{max}+\sigma$ and $\mu_0H_{Cirr} = \mu_0H_{max}-A_s$. This fitting procedure allows for the extraction of the $\mu_0H_{c2}$ and $\mu_0H_{Cirr}$ values from the resistance curves measured at various temperatures, in agreement with the described geometrical procedure. By applying both methods, we ensure consistency and accuracy in determining these critical magnetic field values. Fig. 3 presents the derivative of $R$ with respect to the magnetic field for both samples, measured at the three facilities alongside the $\mu_0H_{c2}$ (superior black circles) and $\mu_0H_{Cirr}$ (inferior black circles). The data from the three facilities are consistent, highlighting the robustness of the observed phenomena across different experimental setups.

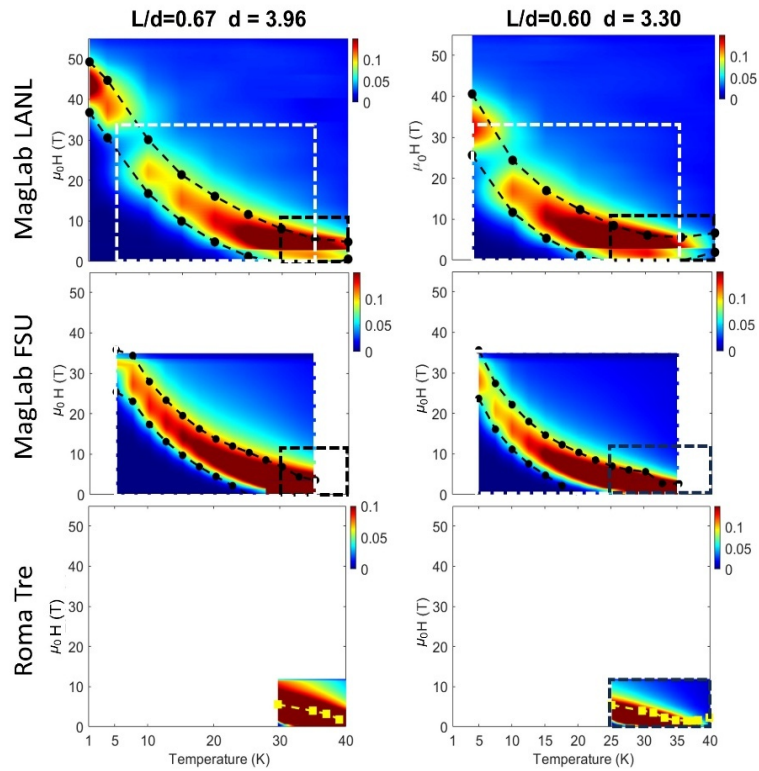

**Figure 3** Color plot of $dR/d(\mu_0H)$ as a function of $\mu_0H$ and $T$, for the first sample. with $L/d$=0.67 and $d$=3.96 nm and the second sample with $L/d$=0.6 and $d$=3.3 nm measured in the three facilities: MagLab at LANL, MagLab at FSU and at Roma Tre University. The specific $T$ - $\mu_0H$ range covered in each facility is well depicted by using the same range in the three facilities.

Further detailed measurements were performed at the Pulsed Field Facility at Los Alamos Laboratory, on a third sample with $d$=5.28 nm and $L/d$ = 0.625 using external perpendicular magnetic field, $\mu_0H_{//c}$.

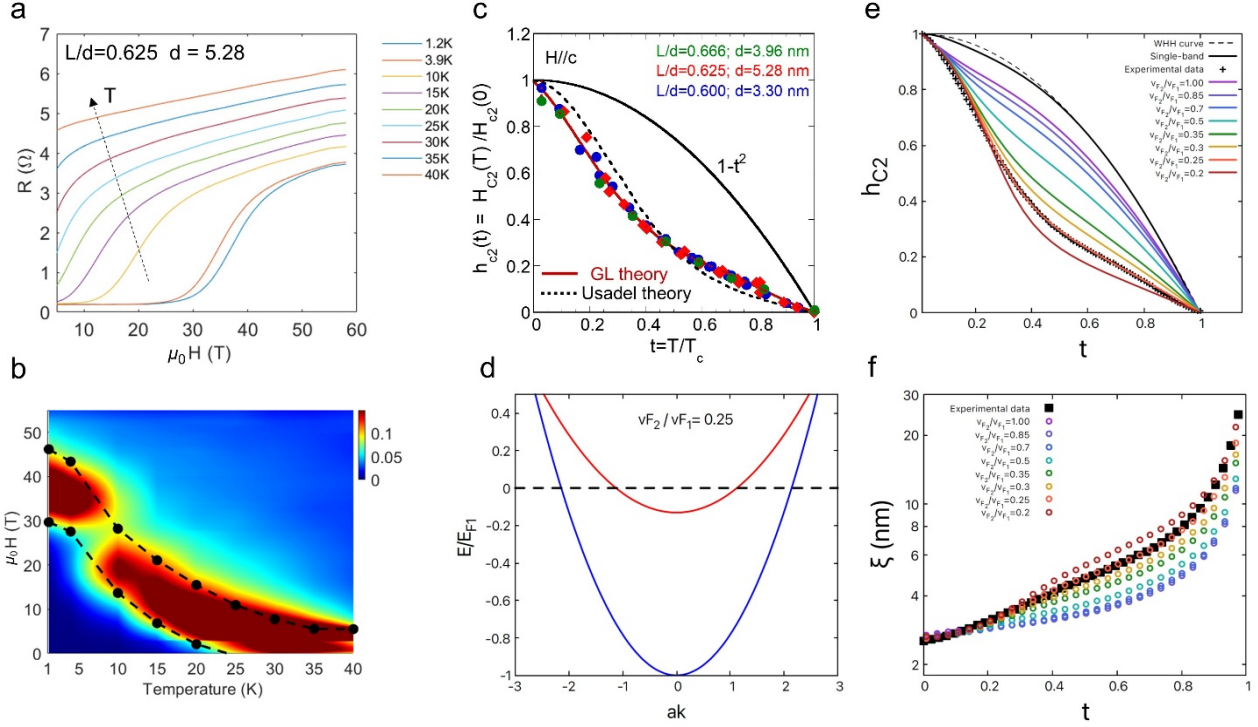

**Figure 4 (a)** Resistance measured in the sample with $L/d$ =0.62 and $d$=5.28 nm, recorded under pulsed high magnetic fields at the MagLab. LANL. **(b)** Color plot $dR/d(\mu_0 H)$ as a function of $\mu_0 H$ and $T$ from the traces in the (a) panel. **(c)** Rescaled curves $h_{c2}(t)$, where $h_{c2}=H_{c2}(T)/H(T = 0$ K$)$, as a function of $t=T/T_c$ for the two AHTS samples in Fig.2 and the third sample in panels a) and b) showing that they collapse on a unique line. The experimental concave $h_{c2}(t)$ curve in panel **(c)** is compared with the theoretical $h_{c2}(t)$ for two band superconductors: i) the dashed black line calculated using Usadel theory with diffusivity ratio η=0.05 in Fig.1 from reference [23] and ii) the solid red line calculated using Ginzburg-Landau theory [42] for the electronic structure of the AHTS samples [16], which is composed of a *deep band* 1 (blue line) and a *shallow band* 2 (red line) having a ratio of the Fermi velocities $v_{F2}/v_{F1} = 0.25$, shown in panel **(d)**. Dashed black line denotes the chemical potential of the system [16,47]. **(e)** Theoretical predictions for the normalized upper critical magnetic field $h_{c2}(t)$ (lines) for two-band superconductors with different Fermi velocity ratios $v_{F2}/v_{F1}$ in each of the two bands compared with the experimental data (crosses). Conventional cases corresponding to a convex curvature of $h_{c2}(t)$ for the BCS theory (dashed line) and the theoretical calculation for a single-band system (solid black line) are reported [23]. Experimental concave curvature of $h_{c2}(t)$ with a kink at $t$=0.4 between the high and low temperature regimes provides compelling experimental evidence for two-gap superconductivity in our AHTS heterostructure tuned at the maximum of the Fano-Feshbach resonance where the ratio of the Fermi velocities between the shallow and deep bands is 0.25. **(f)** Theoretical prediction for the coherence length (open circles) for two-gap artificial superconductors with different Fermi velocity ratios in both bands, the experimental coherence length is represented by the black squares.

Fig. 4 shows the resistance $R$ (panel a) and its derivative with respect to field (panel b) as a function of $\mu_0 H_{//c}$ from the third sample with characteristics $d$=5.28 nm and $L/d$ =0.62, recorded by pulsed high magnetic field at the MagLab LANL. Panel (c) presents the rescaled curves $h_{2c}(t) = H_{c2}(T)/H_{c2}(T=0)$ as a function of $t = T/T_c$, of the three different studied samples which collapse onto a unique line. This rescaling further supports the anomalous concavity in the upper critical magnetic field $h_{c2}$ as a function of *T*, deviating from the convex behavior typically expected in single-band systems. This experimental anomaly of the reduced upper critical magnetic field is a strong indication for superconductivity with two partial condensates associated with two electronic bands in the AHTS superlattices in fact we show that the experimental curve is in agreement with the predicted curves of the theoretical approach of Edge and Balatsky using the Usadel theory [23] and with our curve obtained using the Ginzburg-Landau (GL) theory for a two-band superconductivity model [41-46]. Our approach using the GL theory has been based on the nanomaterial design predictions of the coexistence of first large and second small corrugated cylindrical Fermi surfaces with negative helicity of two quantum subbands where the Fermi velocities ratio is controlled by *L*/*d* [14-16]. We neglect the minor role of the Fermi surface spots with positive helicity due to spin-orbit coupling. For the theoretical interpretation of the presented data, we consider two coupled condensates coexisting in the system in the model two-band system of interacting fermions in 2D which is composed of a first deep band with a high Fermi velocity $v_{F1}$, and a second shallow band with a low Fermi velocity $v_{F2}$ in the proximity of a van-Hove singularity and a Lifshitz electronic topological transition for the appearing of the second small Fermi surface [14,16].

The simple two-band electronic structure model at the top of the superconducting dome shown in Fig.4d, where red and blue continuous lines depict the shallow and deep bands, respectively, grabs the key physical features of the complex electronic structure [14.16] needed for Ginzburg-Landau calculation of the temperature dependent $h_{c2}(t)$ discussed in the *End Matter*. In Fig.4e we show the theoretical upper critical magnetic field $h_{c2}(t)$ for different ratios of the Fermi velocities in the two-bands. We recover the BCS behavior $1-t^2$ (dashed line) for the case where $v_{F2}/v_{F1}$=1. We obtain a good agreement between the theoretical and the experimental $h_{c2}(t)$ for samples at the top of the superconducting dome for *L*=*d*=2/3 with $v_{F2}/v_{F1}$≈0.25. This result is in agreement with the results of Edge and Balatsky [23] using the Usadel theory for the calculation of the upward curvature $h_{c2}(t)$ in interface multiband superconductivity with diffusivities ratio $D_2/D_1$= 0.05 plotted as a dashed black line in Fig.4c. In fact, the ratio of the diffusivity *D* in the two bands is proportional to the square of the Fermi velocities at constant mean free time in the two bands, $\frac{D_2}{D_1} \approx (v_{F2}/v_{F1})^2$ , giving $v_{F2}/v_{F1} = 0.23$.

The evolution of the theoretical coherence length for the different Fermi velocity ratios in the two bands is shown in Figure 4f. The experimental coherence length has been obtained from the upper critical field through the relationship $\mu_0 H_{c2}(T) = \phi_0/(2\pi\xi^2(T))$. Akin to the case of $H_{c2}$, the coherence length exhibits a kink in the region around *t* = 0.4, that becomes more pronounced when the ratio of Fermi velocities becomes progressively smaller and we have the agreement between experimental and theoretical curves for $v_{F2}/v_{F1}$ =0.25.

The low $v_{F2}/v_{F1}$ ratio indicates that a first condensate in the BCS limit with high Fermi velocity resonates with a second condensate with a low Fermi velocity in the BEC-BCS crossover, therefore the experiment provides at the optimal ratio of the Fermi velocities in the two bands for the amplification of $T_c$ at a Fano-Feshbach resonance [17].

*Conclusions -*

The comprehensive study of the superconducting state of the MIMI superlattices, as reported in Ref. [16,17], has confirmed that the observed phenomenology of these superlattices is consistent with a two-band electronic structure and two-gap superconductivity where the band with low Fermi velocity is tuned at a Fano-Feshbach resonance near the Lifshitz transition, causing a sizable amplification of the superconducting critical temperature and gap energies. Regarding the pairing properties, the MIMI phenomenology, as described by Valletta *et al*. [14-16], shows the coexistence of two coupled superconducting condensates one in a BCS-BEC crossover regime, associated with the second upper shallow band, and the other in the BCS regime associated with the first lower deep band. These coupling regimes are associated with a high energy of the bosonic mediator that is of the order of the optical phonon $\omega_0 \approx 60$ meV [47], thus providing the pairing cut-off and yielding $T_c \approx 43$ K in doped $La_2CuO_4$. This allows all pairing channels to be in the intermediate or weak-coupling regime.

By adopting the resonant and multi-condensate scenario discussed above, we have successfully determined the low-temperature superconducting properties of the MIMI superlattices, including the values of the superconducting gaps, upper critical magnetic fields, and superconducting critical temperatures. Through this approach, we extracted the microscopic parameters of the two-band structure and the pairing interactions. This serves as the foundation for the theoretical investigation of the temperature dependence of the upper critical magnetic field and the phase coherence length hence allowing for a quantitative comparison with the experimental data. We demonstrated that the key microscopic factors determining the peculiar shape of $H_{c2}(T)$ are the ratio of the Fermi velocities in the two bands, and the interplay between intraband couplings and the pair-exchange term configuring a coupled, but still distinct, two-gap superconducting condensate. It is impossible to reproduce the well-formed upward concavity of $h_{c2}(t)$ reported in this work with either a single band model or with similar Fermi velocities for both bands. Finally, we have focused on nanoscale AHTS heterostructures with lattice geometry factor, that is the ratio $L/d$ near 2/3, at the top of the superconductivity dome. These samples are in the proximity of a Fano-Feshbach resonance between the two-gap coupled condensates with optimal ratio of the two Fermi velocities in two-bands. Our theoretical approach predicts that the temperature dependence of $h_{c2}(t)$ approaches the standard convex behavior inherent to single-band systems if both velocities are very similar. In contrast, here we provide experimental evidence for the upward concavity with optimal ratio between both Fermi velocities around 0.25 near the Fano-Feshbach resonance [14-16].

The implications of this study are profound, as they not only enhance our understanding of the fundamental mechanism of high-Tc superconductivity in La-based cuprates but also pave the way for future research into the design and optimization of new high-performance superconducting materials emerging from key multicomponent and resonant quantum phenomena.

**End Matter - Material and Methods**

**The synthesis of artificial high-T$_c$ superlattices (AHTS)** was achieved using molecular beam epitaxy (MBE). These superlattices, composed of normal metal $La_{1.55}Sr_{0.45}CuO_4$ (LSCO) alternating with superconducting space-charge layers in $La_2CuO_4$ (LCO) thin films, were synthesized via an ozone-assisted MBE method (DCA Instruments Oy, Turku, Finland) on $LaSrAlO_4$ (001) substrates. The compressive strain for $La_2CuO_4$ on $LaSrAlO_4$ is +1.4%. The growth of the superlattices was monitored using in situ reflection high-energy electron diffraction (RHEED). This method is characterized by the sequential deposition of single atomic layers and the minimal kinetic energy of impinging atoms (approximately 0.1 eV). The substrate temperature ($T$s), as measured by a radiation pyrometer, was maintained at 650 °C, with a chamber pressure of approximately $1.5 \times 10^{-5}$ Torr, consisting of mixed ozone, atomic, and molecular oxygen. Upon completion of the procedure, the samples were cooled down to $T_s$ = 200 °C. Afterwards the delivery of ozone was shut down and samples were cooled down in vacuum to remove the interstitial oxygen in $La_2CuO_4$ layers and to make these layers stoichiometric. The high-quality crystal superlattice structure of the AHTS was confirmed by X-ray diffraction at the Trieste Elettra synchrotron radiation facility.

**Magneto resistance experiments** were performed at the National High Magnetic Field Laboratory (NHMFL) of Florida State University in Tallahassee to measure the upper magnetic fields in artificial high-T$_c$ superlattices as a function of the temperature. High magnetic fields were provided by a Bitter resistive magnet capable of generating fields up to $\mu_0 H$ =35 T. This system was coupled to a variable temperature insert able to ramp, or stabilize, the temperature within the range 1.35 K ≤ $T$ ≤ 300 K via calibrated Cernox thermometers. A single axis rotator, coupled to a Hall probe, was used to orient the substrates along either the *c*-axis or the *ab*-plane. Resistance bridges (Lakeshore 370) were used to measure the four-point resistance in a van der Pauw configuration with Au pads evaporated onto the corners of the substrates acting as the electrical contacts. An excitation current of 3 µA was used after checking that it yields identical results to those collected with an excitation of 10 µA."

**Magnetotransport measurements** were performed using a standard four-point contact method at the Pulsed Field Facility of the National High Magnetic Field Laboratory at Los Alamos Laboratory. The samples were mounted on a custom-designed probe compatible with pulsed magnetic fields. Electrical contacts were made using silver epoxy and copper wires. The pulsed magnetic field profile, with a rise time of ~10 ms and a total duration of ~100 ms, was synchronized with data acquisition. Temperature control was achieved using a helium-3 cryostat, allowing measurements down to 0.5 K. Resistance as a function of magnetic field and temperature was recorded using lock-in amplifier, ensuring high signal-to-noise ratio. Typical measurement frequencies of ~64 kHz and sample currents of ~80 mA were utilized. Data were collected during both the upsweep up to 58 Tesla and down-sweep of the magnetic field to account for any hysteretic effects.

**Temperature dependent sheet resistance** at constant magnetic field parallel to the crystallographic c-axis, in the range $0T \leq \mu_0 H \leq 12T$, were performed at Roma Tre University, ramping the magnetic field at 0.2 T/min. Data were acquired using a standard four points contact method with 10 $\mu A$ dc excitation and measuring the voltage with a Keithley 2182a nanovoltmeter in delta mode. Measurements were performed in zero field cooling conditions, after verifying, through comparison with measurements performed in field cooling, that the magnetization effects of the sample on the $R$ measurements were negligible.

**Theoretical methods.** Here we discuss the physical method used to explain the experimental $h_{c2}(t)$ data for the multiband superconductivity in our AHTS with complex multiband electronic structure [14,16] summarized by the pictorial view in **Fig. 5**.

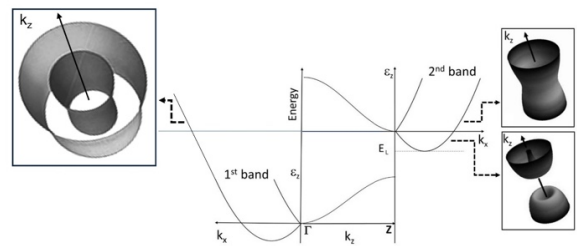

**Figure 5.** The two cylindrical Fermi surfaces for the first and second bands generated by quantum size effect at the top of the superconducting dome. The spin-orbit coupling splits the cylindrical Fermi surface into an internal and external surface with positive and negative helicity, respectively. The Fermi surface of the second band is in the proximity of a van Hove singularity near the electronic topological transition from a torus to a corrugate cylinder.

The pairing interaction among electrons has been approximated by a separable attractive potential $V_{ij}(k,k')$ with an energy cutoff. To calculate the temperature dependence of $\mu_0 H_{c2}$, we first evaluate the in-plane coherence length $\xi$ which is related to $\mu_0 H_{c2}$ in the direction perpendicular to the plane through the relation: $\mu_0 H_{c2} = \Phi_0/2\pi\xi^2(T)$. Regarding the phase coherence length, in two-band systems two characteristic length scales in the spatial behavior of the superconducting fluctuations and order parameter are expected. When the pair-exchange interactions are not present, these two lengths reduce to the coherence lengths of the two partial condensates. When the pair-exchange interactions are not zero, one has to deal with coupled condensates, and these length scales cannot be attributed to the single bands involved, describing instead the collective features of the whole two-component superconducting condensate. The soft, or critical, coherence length $\xi_s$ diverges at the transition temperature and governs the inter-pair phase coherence, while the rigid, or non-critical, coherence length $\xi_r$ remains always finite. In this work, we evaluate the soft coherence length $\xi_s \approx \xi$ using the Ginzburg-Landau (GL) approach [42,43], which allows us to determine $\mu_0 H_{c2}$. We define the dimensionless couplings $\lambda_{ij}$ as $\lambda_{ij} = V_{ij}^0 N_j$, where $N_j = m_j^*/2\pi\hbar^2$ is the 2D density of states per spin and per unit area of the j-th band and $V_{ij}^0$ are the matrix elements of the separable attractive potential. The energy cutoff of the effective pairing interaction is 60 meV. We choose the coupling constants to obtain a coherence length at $T \approx 0$K of $\approx 2.6$ nm for all the cases so that $\mu_0 H_{c2}(T \approx 0) \approx 48$ T, while the mean-field superconducting critical temperature results $T_c \approx 45$ K. We use for this purpose $\lambda_{11}= 0.025$, $\lambda_{22}= 0.308$, $\lambda_{12} = 0.174$ and $\lambda_{21} = 0.083$. We consider an effective mass ratio $m_2^*/m_1^* = 2.1$. With this choice of parameters, by solving the mean-field self-consistent gap equations we have obtained a zero-temperature gap ratio $\Delta_2/\Delta_1 \approx 2$.

*Acknowledgements* - We thank the Superstripes onlus and the CNR project PdGP/GePro 2024-2026 for supporting this work. L.B. is supported by the US DOE-BES through award DE-SC0002613. The National High Magnetic Field Laboratory is supported by the National Science Foundation through NSF/DMR-2128556*, the State of Florida, and the U.S. Department of Energy.

**References**
1. A. Ohtomo, H. Y. Hwang, A high-mobility electron gas at the LaAlO$_3$/SrTiO$_3$ heterointerface, *Nature*, **427**, 423-426 (2004).
2. N. Reyren, S. Thiel, A. D. Caviglia, et al., Superconducting interfaces between insulating oxides, *Science*, **317**(5842), 1196-1199 (2007).

3.  A. Bianconi, D. Innocenti, A. Valletta, A. Perali, Shape Resonances in superconducting gaps in a 2DEG at oxide-oxide interface, *J. Phys.: Conf. Ser.* **529** (2014).

4.  D. Valentinis, S. Gariglio, A. Fête, J.M. Triscone, C. Berthod, D. Van Der Marel, Modulation of the superconducting critical temperature due to quantum confinement at the $LaAlO_3/SrTiO_3$ interface, *Phys. Rev. B* **96**, 094518 (2017)

5.  G. Singh, G. Venditti, G. Saiz, et al., Two-gap s±-wave superconductivity at an oxide interface, *Physical Review B*, **105**(6), 064512 (2022).

6.  A. Bianconi, A. Valletta, A. Perali, N.L. Saini, Superconductivity of a striped phase at the atomic limit, *Physica C Supercond.* **296**, 269–280 (1998)

7.  A. Bianconi, Feshbach shape resonance in multiband superconductivity in heterostructures. *J. Supercond.* **18**, 625–636 (2005).

8.  M. Cariglia, A. Vargas-Paredes, M.M. Doria, A. Bianconi, M.V. Milošević, A. Perali, Shape-resonant superconductivity in nanofilms: from weak to strong coupling, *J. Supercond. Nov. Magn.* **29**, 3081–3086 (2016)

9.  Y. Chen, A.A. Shanenko, A. Perali, F.M. Peeters, Superconducting nanofilms: molecule-like pairing induced by quantum confinement, *Journal of Physics: Condensed Matter* **24**, 185701 (2012).

10. L. Salasnich, A.A. Shanenko, A. Vagov, J.A. Aguiar, A. Perali, A Screening of pair fluctuations in superconductors with coupled shallow and deep bands: A route to higher-temperature superconductivity, *Phys. Rev. B*, **100**, 064510 (2019).

11. F. Baiutti, G. Logvenov, G. Gregori, et al., High-temperature superconductivity in space-charge regions of lanthanum cuprate induced by two-dimensional doping, *Nature Communications*, **6**(1), 8586 (2015)

12. Y. E. Suyolcu, G. Christiani, P. A. Van Aken, and G. Logvenov, Design of complex oxide interfaces by oxide molecular beam epitaxy, *Journal of Superconductivity and Novel Magnetism*, **33**, 107-120 (2020).

13. D. Mondal, S.R. Mahapatra, A.M. Derrico, R.K. Rai, J.R. Paudel, C. Schlueter, A. Gloskovskii, R. Banerjee, A. Hariki, F.M.F. DeGroot, et al., Modulation-doping a correlated electron insulator. *Nat. Commun.*, **14**, 6210 (2023).

14. M.V. Mazziotti, A. Bianconi, R. Raimondi, G. Campi, A. Valletta, Spin–orbit coupling controlling the superconducting dome of artificial superlattices of quantum wells. *J. Appl. Phys.,* **132**, 193908 (2022)

15. G. Logvenov, N. Bonmassar, G. Christiani, G. Campi, A. Valletta, A. Bianconi, The superconducting dome in artificial high-$T_c$ superlattices tuned at the Fano–Feshbach resonance by quantum design. *Condens. Matter,* **8**, 78 (2023).

16. A. Valletta, A. Bianconi, A. Perali, G. Logvenov, G. Campi, High-$T_c$ superconducting dome in artificial heterostructures made of nanoscale quantum building blocks. *Phys. Rev. B* **110**, 184510 (2024)

17. G. Campi, G. Logvenov, S. Caprara, A. Valletta, A. Bianconi, Kondo versus Fano in superconducting artificial high-$T_c$ heterostructures. *Condensed Matter* **9**, 43 (2024).

18. K. Ochi, H. Tajima, K. Iida, H. Aoki, Resonant pair-exchange scattering and BCS-BEC crossover in a system composed of dispersive and heavy incipient bands: A Feshbach analogy. *Phys. Rev. Res.,* **4**, 013032.e26 (2022).

19. H. Tajima, H. Aoki, A. Perali, A. Bianconi, Emergent Fano-Feshbach resonance in two-band superconductors with an incipient quasiflat band: enhanced critical temperature evading particle-hole fluctuations. *Physical Review B*, **10**, L140504 (2024).

20. N.R. Werthamer, E. Helfand, P.C. Hohenberg, Temperature and purity dependence of the superconducting critical field, $H_{c2}$. III. Electron Spin and Spin-Orbit Effects. *Physical Review*, **147**, 295–302 (1966).

21. M. H. Jung, M. Jaime, A. H. Lacerda, G. S. Boebinger, et al., Anisotropic superconductivity in epitaxial $MgB_2$ films. *Chemical Physics Letters,* **343**(5-6), 447–451. (2001).

22. M. E. Palistrant, V. A. Ursu, Thermodynamic and magnetic properties of superconductors with anisotropic energy spectrum, $MgB_2$. *Journal of Superconductivity and Novel Magnetism*, **21**(3), 171–176 (2008).

23. J. M. Edge, A. V. Balatsky, Upper critical field as a probe for multiband superconductivity in bulk and interfacial STO. *Journal of Superconductivity and Novel Magnetism*, **28**, 2373-2384 (2015).

24. R.M. Fernandes, J.T. Haraldsen, P. Wolfle, A.V. Balatsky, Two band superconductivity in doped $SrTiO_3$ films and interfaces, Phys. Rev. B **87**(1), 014510 (2013).

25. A. Bussmann-Holder, A. R. Bishop, A. Simon, $SrTiO_3$: From Quantum Paraelectric to Superconducting, *Ferroelectrics,* **400**(1), 19-26 (2010).

26. Y. J. Jo, J. Jaroszynski, A. Yamamoto, A. Gurevich, S. C. Riggs, G. S. Boebinger, … L. Balicas, High-field phase-diagram of Fe arsenide superconductors, *Physica C: Superconductivity*, **469**(9-12), 566–574 (2009).

27. M. Kano, Y. Kohama, D. Graf, F. Balakirev, *et al.,* Anisotropy of the upper critical field in a Co-doped $BaFe_2As_2$ single crystal, *Journal of the Physical Society of Japan,* **78**(8), 084719 (2009).

28. T. Tamegai, Y. Nakajima, T. Nakagawa, G. J. Li, H. Harima, Two-gap superconductivity in $R_2Fe_3Si_5$ (R= Lu and Sc), In *Journal of Physics: Conference Series* **150**(5), 052264 (2009).

29. F. Hunte, J. Jaroszynski, A. Gurevich, D. C. Larbalestier, et al., Two-band superconductivity in $LaFeAsO_{0.89}F_{0.11}$ at very high magnetic fields, *Nature*, **453**(7197), 903-905 (2008).

30. M. Bristow, A. Gower, J. C. A. Prentice, M.D. Watson, *et al.*, Multiband description of the upper critical field of bulk FeSe. *Physical Review B,* **108**(18), 184507 (2023).

31. A. Vashist, B. R. Satapathy, H. Silotia, Y. Singh, S. Chakraverty, Multigap superconductivity with non-trivial topology in a Dirac semimetal PdTe, *arXiv preprint* arXiv:2408.06424 (2024).

32. T. B. Charikova, N. G. Shelushinina, G. I. Harus, D. S. Petukhov, V. N. Neverov, A. A. Ivanov, Upper critical field in electron-doped cuprate superconductor $Nd_{2-x}Ce_xCuO_{4+\delta}$: Two-gap model, *Physica C: Superconductivity*, **488**, 25–29 (2013).

33. M. E. Palistrant, The upper critical field $H_{c2}$ in advanced superconductors with anisotropic energy spectrum, *Journal of Superconductivity and Novel Magnetism,* **23**(8), 1427–1442(2010).

34. A. Gurevich, Enhancement of the upper critical field by nonmagnetic impurities in dirty two-gap superconductors, *Physical Review B*, **67**(18), 184515 (2003).

35. A. Gurevich, Limits of the upper critical field in dirty two-gap superconductors, *Physica C: Superconductivity*, **456**, 160–169 (2007).

36. H. S. Lee, M. Bartkowiak, J. H. Park et al. (2009). Effects of two gaps and paramagnetic pair breaking on the upper critical field of SmFeAsO$_{0.85}$ and SmFeAsO$_{0.8}$F$_{0.2}$ single crystals. *Physical Review B—Condensed Matter and Materials Physics*, **80**(14), 144512.

37. T. Tamegai, Y. Nakajima, T. Nakagawa, G. J. Li, H. Harima, Two-gap superconductivity in R$_2$Fe$_3$Si$_5$ (R= Lu and Sc). *Journal of Physics: Conference Series* **150**(5) 052264 (2009).

38. Jl. Zhang, L. Jiao, Y. Chen, et al., Universal behavior of the upper critical field in iron-based superconductors, *Front. Phys*. **6**, 463–473 (2011).

39. A. I. Coldea, D. Braithwaite, A. Carrington, Iron-based superconductors in high magnetic fields. *Comptes Rendus Physique,* **14**(1), 94–105 (2013).

40. Y. Liu, R. Wang, Z. Wen, J. Shu, Y. Cui, Y. Chen, Y. Zhao, The upper critical field, anisotropy, and critical current density of superconducting LaO$_{1-x}$BiS$_2$ with x=0.07, *Journal of Superconductivity and Novel Magnetism,* **34**, 1157-1163 (2021).

41. İ. N. Askerzade, A. Gencer, and N. Güçlü. On the Ginzburg-Landau analysis of the upper critical field H$_{c2}$ in MgB$_2$. *Superconductor science and technology* 15.2 (2002): L13. **DOI** 10.1088/0953-2048/15/2/102

42. G. Litak, T. Ord, K. Rago, A. Vargunin, Coherence lengths for superconductivity in the two-orbital negative-U Hubbard model, *Acta Physica Polonica A*. **121**, 747 (2012).

43. G. Midei, A. Perali, Sweeping across the BCS-BEC crossover, reentrant, and hidden quantum phase transitions in two-band superconductors by tuning the valence and conduction bands, *Phys. Rev. B* **107**, 184501 (2023).

44. A. Guidini, A. Perali, Band-edge BCS–BEC crossover in a two-band superconductor: physical properties and detection parameters, *Superconductor Science and Technology* **27**, 124002 (2014).

45. L. Komendová, Y. Chen, A.A. Shanenko, M.V. Milošević, F.M. Peeters, Two-band superconductors: hidden criticality deep in the superconducting state, *Phys. Rev. Lett.* **108**, 207002 (2012).

46. N.H. Aase, C. S. Johnsen, and A. Sudbø, Constrained weak-coupling superconductivity in multiband superconductors, *Phys. Rev. B* **108**, 024509 (2023).

47. T. Egami, Electron-phonon coupling in high-T$_C$ superconductors, in Müller, K.A., Bussmann-Holder, A. (eds) Superconductivity in Complex Systems. *Structure and Bonding*, vol 114, p. 268-283 Springer, Berlin, Heidelberg.